\documentclass[aps,prl,twocolumn,showpacs,amsmath,amssymb,10pt]{revtex4-1} 
\usepackage{graphicx,subfig,xcolor}
\usepackage{manfnt,pifont,dsfont}

\newcommand\del{\partial}

\newcommand{\be}{\begin{equation}}
\newcommand{\ee}{\end{equation}}
\newcommand{\bea}{\begin{eqnarray}}
\newcommand{\eea}{\end{eqnarray}}
\newcommand\restr[2]{{\left.\kern-\nulldelimiterspace#1\vphantom{\big|}\right|_{#2}}}

\newcommand{\reef}[1]{(\ref{#1})}

\newcommand{\beq}{\begin{equation}} 
\newcommand{\eeq}{\end{equation}}
\def\del {\partial} 
 
\def\bZ {\mathbb{Z}}

\def\calO {{\cal O}}
\def\calR {{\cal R}}

\def\bZ {\mathbb{Z}}

\def\geq{\geqslant}

\def \del{\partial}

\newcommand{\mysigma}{s}
\newcommand{\lr}{{\rm LRFP}}
\newcommand{\sr}{{\rm SRFP}}

\begin{document}
\title{Long-range critical exponents near the short-range crossover 
}
\author{Connor Behan$^a$, Leonardo Rastelli$^a$, Slava Rychkov$^{b, c}$, and Bernardo Zan$^{d,b}$ } 
\affiliation{
$^a$  C.~N.~Yang Institute for Theoretical Physics, SUNY, Stony Brook, NY 11794-3840, USA\\
$^b$  CERN, Theoretical Physics Department, 1211 Geneva 23, Switzerland\\
$^c$  Laboratoire de Physique Th\'eorique de l'\'Ecole Normale Sup\'erieure, PSL Research University, CNRS, Sorbonne Universit\'es, UPMC Univ.\,Paris 06, 24 rue Lhomond, 75231 Paris Cedex 05, France\\
$^d$  Institut de Th\'eorie des Ph\'enom\`enes Physiques, EPFL, CH-1015 Lausanne, Switzerland
}

\begin{abstract}
The  $d$-dimensional  long-range Ising model, defined by spin-spin interactions decaying with the distance as the power $1/r^{d+s}$, admits a second order phase transition 
 with continuously varying critical exponents. At $s = s_*$, the phase transition crosses over to the usual short-range universality class.
The standard field-theoretic description of this family of models   is strongly coupled at the {crossover}. We find a new  description, which is instead weakly coupled near the crossover, and use it to compute critical exponents. The existence of two complementary UV descriptions of the same long-range fixed point provides a novel example of infrared duality.
\end{abstract}
\pacs{   Preprint numbers: YITP-SB-17-12, CERN-TH-2017-058}
\maketitle
\nopagebreak
{\bf Introduction.}
Statistical models with long-range interactions exhibit rich critical behavior with continuously varying exponents. Here we will focus for definiteness on the long-range Ising model (LRI) in $d=2$ and $d=3$ dimensions \cite{note1}, with ferromagnetic interaction between spins decaying as a power of their distance as $1/r^{d+\mysigma}$, where  $\mysigma >0$ for the thermodynamic limit to be well defined \cite{note2}. 
 There are  three critical regimes depending on the value of the microscopic parameter: (i) the mean-field regime for $\mysigma<d/2$, (ii) the intermediate regime for $d/2<\mysigma<\mysigma_*$, and (iii) the short-range regime for $\mysigma>\mysigma_*$. Our primary goal will be to elucidate the long-range to short-range crossover at $\mysigma=\mysigma_*$, but let us first briefly review all three regimes.  

The most convenient way to study the long-distance behavior is to replace the lattice model with a continuum field theory in the same universality class. Apart from the usual quadratic and quartic terms that are both local, the action includes a gaussian non-local term, with a negative sign for the considered ferromagnetic case \cite{Note101}:
 \beq 
 \label{standardflow}
 S= - \int d^dx\, d^dy\,  \frac{ \phi(x)\phi(y)}{|x-y|^{d+\mysigma} }+ \int d^d x [t\,\phi(x)^2 + g\, \phi(x)^4]\,. 
  \eeq
The non-local term by itself describes a mean field theory in which $\phi$ has the dimension $[\phi]_{\rm UV}=(d-\mysigma)/2$ \cite{note3}. The quadratic term is always relevant and the transition is reached by tuning its coefficient $t$ to zero \cite{Note100}. The quartic term is irrelevant for $\mysigma<d/2$, explaining why the transition becomes mean-field below this value.  On the other hand, for $\mysigma>d/2$  
the quartic interaction
induces a nontrivial RG flow. In the regime (ii) $d/2<\mysigma<\mysigma_*$, this flow ends in an interacting \emph{long-range} fixed point (LRFP). General composite operators 
such as $\phi^2$ acquire nontrivial anomalous dimensions, as befits an interacting fixed point.
However, the dimension of $\phi$ is controlled by a non-local term and is therefore not renormalized: $[\phi]_{\lr}=[\phi]_{\rm UV}=(d-\mysigma)/2$ at the LRFP.
Finally, the long-range to short-range crossover is expected to happen \cite{Sak} when $[\phi]_{ \lr}$, decreasing with $\mysigma$, reaches the short-range Ising fixed point (SRFP) dimension $[\phi]_{\sr}$. In other words, the dimension of $\phi$ is continuous at the crossover, fixing $\mysigma_*=d - 2  [\phi]_{\sr} \equiv 2-\eta_{\sr}$. 

\begin{figure}[h!tbp]
\begin{center}
\subfloat[][]{\includegraphics[scale=0.6]{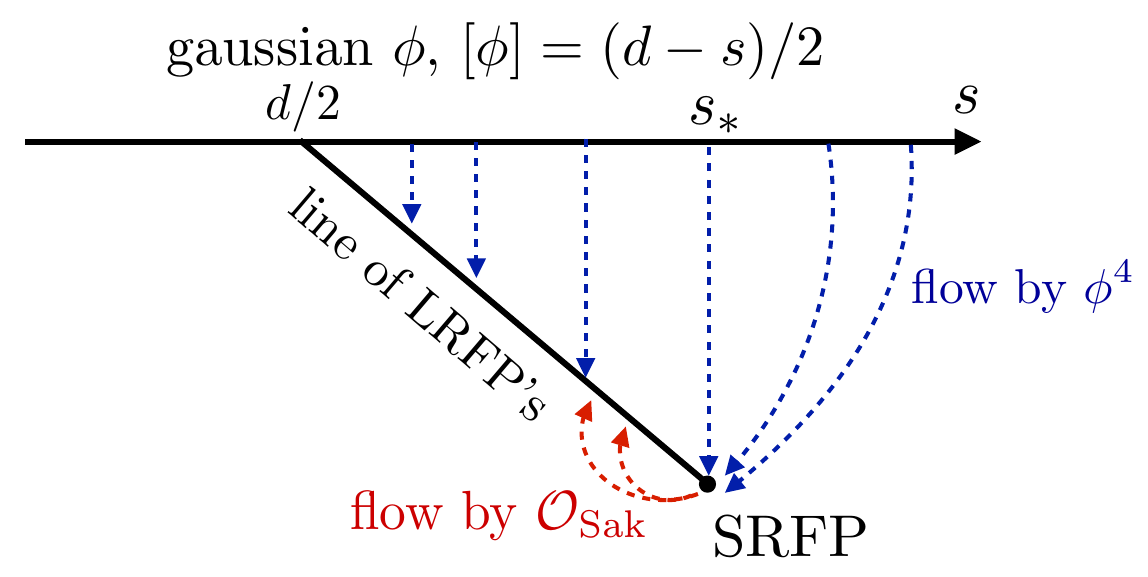}} \qquad
\\
\subfloat[][]{\includegraphics[scale=0.6]{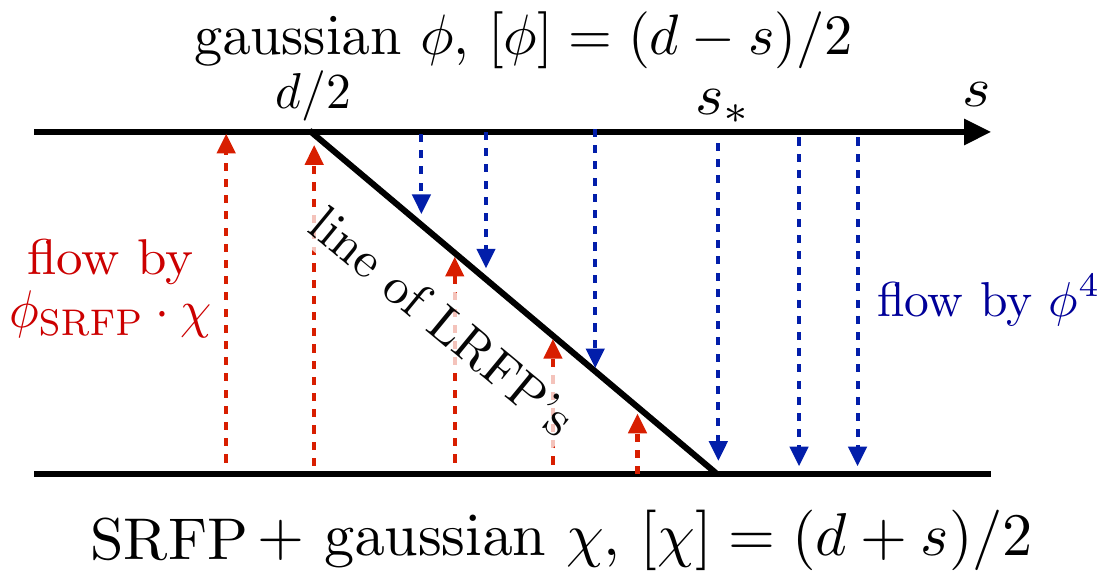}}
\caption{RG flows corresponding to (a) the standard picture and (b) our proposal.}
\label{fig-standard}
\end{center}
\end{figure}

The  picture that we have just reviewed  is considered standard since the original work by Fisher et al. \cite{Fisher} and its refinement by Sak \cite{Sak, Sak2}, but 
while the crossover from the mean-field to the intermediate regime is well understood, the same cannot be said of the long-range to short-range crossover. For $\mysigma$ slightly above $d/2$, the quartic interaction is weakly relevant and one can study the flow in perturbation theory,
computing physical quantities in a systematic expansion in $\epsilon = 2 \mysigma - d$. By contrast, a perturbative description of the long-range to short-range crossover is lacking at present. The non-local perturbation
\beq 
\calO_{\rm Sak}=\int d^dx\, d^dy\, \frac{ \sigma(x)\sigma(y)} {|x-y|^{d+\mysigma}} \,,
\eeq 
where $\sigma\equiv \phi_{\sr}$ is the SRFP spin field, has been proposed \cite{Sak2, Cardy} as a way to analyze the short-range fixed point stability. The critical $\mysigma=\mysigma_*$ is precisely where this perturbation crosses from relevant to irrelevant \cite{Sak2, Cardy}.
For $\mysigma<\mysigma_*$, the flow from the short-range fixed point perturbed by $\calO_{\rm Sak}$ should end in the long-range fixed point. The RG flow diagram which summarizes the standard picture is shown in Fig.~\ref{fig-standard}(a). If $\mysigma$ is just slightly below $\mysigma_*$, it should in principle be possible to study the flow perturbatively because the perturbation $\calO_{\rm Sak}$ is weakly relevant. 
However, it is unclear how the rules of conformal perturbation theory should be adapted to this {{non-local}} case.

One may dismiss this lack of computability as a technical problem, but there is a related conceptual puzzle. If the crossover is continuous, the entire operator spectrum should be continuous, not just $\phi$. Consider however $\phi^3$. The non-local equation of motion that follows from the action (\ref{standardflow}) relates $\phi$ and $\phi^3$,
 implying that at the IR fixed point their dimensions must obey  \cite{LRI} the ``shadow relation''  
\beq \label{shadowphi}
[\phi]_{\rm LRFP} + [\phi^3]_{\rm LRFP} = d\,. 
\eeq 
This means that at the crossover there should be a $\mathbb{Z}_2$ odd operator of dimension $d - [\phi]_{\rm SRFP}$, in contradiction with the well-established fact
that the short-range Ising fixed point contains a {\it single} relevant $\mathbb{Z}_2$ odd scalar. A similar puzzle arises for the stress tensor operator. 
 A local conserved stress tensor $T_{\mu\nu}$ exists in the short-range fixed point. As we move to the long-range regime, this operator is expected to acquire an anomalous dimension so that it is no longer conserved. The divergence $V_{\nu}=\del^\mu T_{\mu\nu}$ is thus a nontrivial local operator at the long-range fixed point, but where did it come from? 
The short-range Ising fixed point does {\it not} contain a vector conformal primary
 operator of dimension exactly $d+1$ that could play the role of $V_{\mu}$ \cite{noteRecomb}.
 
{\bf Our proposal.} The need to resolve the above difficulties leads us to suggest that the crossover happens not just to the short-range fixed point, but rather to the short-range fixed point \emph{plus} a decoupled gaussian field $\chi$, a theory we dub ``SRFP+$\chi$". In our proposal, the flow from SRFP+$\chi$ to the long-range fixed point is driven by the perturbation
\beq \label{chiphi}
g_0 \int d^dx\, \calO(x),\quad \calO = \sigma \cdot \chi \,.
\eeq
The standard picture is recovered by integrating out $\chi$, which  generates precisely the {{non-local}} perturbation $\calO_{\rm Sak}$ discussed above \cite{noteSign}. This fixes $\chi$ to have the dimension $[\chi]=(d+\mysigma)/2$, so that
\beq
[\calO]=[\chi]+[\sigma]=d-\delta,\quad \delta = (s_*-s)/2\,,
\eeq 
crossing from relevant to irrelevant at the same location as before.  We emphasize however that $\chi$ is not just a theoretical construct introduced to represent the {{non-local}} perturbation $\calO_{\rm Sak}$. Rather, it is a physical field, which can be thought of as a remnant of the long-range interactions of the original model \reef{standardflow} \cite{noteCoincides}.

The new RG flow diagram is shown in Fig.~\ref{fig-standard}(b). In the intermediate regime $d/2 < \mysigma < \mysigma_*$, two distinct RG flows have the long-range fixed point as their common IR endpoint: the standard flow (\ref{standardflow}) from the mean field theory, which is weakly coupled near the lower end of the intermediate regime $(\epsilon \to 0$), and our newly proposed flow emanating from the SRFP+$\chi$ theory, which is weakly coupled near the crossover $(\delta \to 0$).  This fits the classic pattern of a field-theoretic infrared duality. 

{\bf Beta-function.} In the rest of this Letter, we will use the flow from SRFP+$\chi$, perturbed by \reef{chiphi}, to obtain a new quantitative understanding of the long-range fixed point near the crossover. A more detailed presentation of our results will appear in \cite{LRI2}. 

As $\calO$ is a local operator, we can employ the standard framework of conformal perturbation theory \cite{Zamolodchikov:1987ti,Brezin:1990sk,Cappelli:1991ke}. To recall briefly, consider the order $n$ perturbative correction to the observable $\calO(\infty)=\lim_{x\to\infty} x^{2\Delta_\calO} \calO(x)$:
\beq
\label{Og^n}
\frac{g_0^n}{n!}\int d^dx_1\ldots d^dx_n \langle \calO(x_1)\ldots\calO(x_n)\,\calO(\infty)\rangle\,.
\eeq
Divergences due to colliding $x_i$ require us {{to}} introduce a regulator which is most easily taken to be a short-distance cutoff $a$. The beta-function is found by demanding that \reef{Og^n} be independent of $a$ when expressed in terms of the renormalized coupling $g = a^\delta g_0$. It is therefore related to the logarithmically divergent part of the integral \cite{noteDiv}.

In our case $\beta(g)= -\delta g + \beta_3g^3 + O(g^5)$, as all even-order contributions vanish by the $\bZ_2\times \bZ_2$ symmetry of the SRFP+$\chi$ theory (independent sign flips of $\sigma$ and $\chi$). Extracting the logarithmic divergence, the coefficient $\beta_3$ is expressed as an integral of the  four-point function in particular kinematics \cite{Zohar}:
\beq
\beta_3=-\frac 1 6 {\rm S}_{d} \int d^d x\, \langle \calO(0) \calO(\hat{e}) \calO(x)\calO(\infty)\rangle\,.
\label{eq:beta}
\eeq
Here $\hat{e}$ is an arbitrary unit-length vector, and ${\rm S}_{d}$ is the unit sphere area. Although this integral is still infinite due to the relevant operators in the $\calO \times \calO$ operator product expansion (OPE), its divergences at $x\to0,\hat e,\infty$ are powerlike and unrelated to the running of $g$. To handle them, we exploit the symmetry between the three channels to write \reef{eq:beta} as an integral over the region $\calR = \{ y : |y| < 1, |y| < |y - \hat{e}| \}$. As explained in \cite{Zohar}, this isolates $y \to 0$ as the single place where we have to subtract pure powers. 

The four-point function of $\calO$ factorizes into the product of the four-point functions of $\sigma$ and $\chi$. The four-point function of $\chi$, being gaussian, is given by the sum of three independent Wick contractions, while the four-point function of $\sigma$ is nontrivial. In $d=2$ it is known in closed form thanks to the exact solution of the corresponding minimal model CFT \cite{BPZ}, while in $d=3$  it is known approximately with high precision from the numerical conformal bootstrap \cite{c-min},\cite{Zohar,lightcone}. For the $d=3$ computation, we have used data from the 3d
Ising spectrum up to dimension $\Delta_* = 8$ \cite{lightcone}.

Performing numerically the steps outlined above, we have found \cite{noteErrors}
\begin{eqnarray}
\beta_3 = 1.2684040(5) & &\qquad(d = 2) \\
\beta_3 = 12.26(3)\qquad & &\qquad(d = 3)\, .
\end{eqnarray}
 The sign of $\beta_3$ was not manifest in the above calculations, since the regulated integrals are not sign-definite. Still, we see that $\beta_3$ is positive in both $d=2$ and $d=3$. This is a basic check of our proposal: as expected, $\calO$ is marginally irrelevant at the crossover. As $\delta>0$, we have an IR fixed point at $g_*^2=\delta/\beta_3+O(\delta^2)$.

{\bf Anomalous dimensions.} To study the renormalization of an operator $\Phi$, we are instructed to consider the order $n$ correction to its two-point function:
\beq
\label{og^n}
\frac{g_0^n}{n!}\int d^dx_1\ldots d^dx_n \langle \Phi(0)\calO(x_1)\ldots\calO(x_n)\,\Phi(\infty)\rangle\,.
\eeq
Using point splitting with the cutoff $a$ once again, logarithmic divergences are absorbed into $Z_\Phi(g, a)$ so that the renormalized operators $\Phi_{\mathrm{R}} \equiv Z_\Phi(g, a)^{-1} \Phi$ have finite correlators. We have $Z_\Phi = 1 + \frac{g^2}{2}B\log\frac{1}{a} + O(g^4)$ as a consequence of the $\bZ_2\times \bZ_2$ symmetry. Equivalently, the anomalous dimension at the fixed point is given by $\gamma_\Phi = -\frac{g_*^2}{2}B + O(g_*^4)$. Isolating the logarithmic divergence of \reef{og^n},
\beq
B = {\rm S}_{d} \int d^d x\, \langle \Phi(0)\calO(x)\calO(\hat{e})\Phi(\infty) \rangle\,.
\label{eq:B}
\eeq
This power divergent expression can be regulated using the same region $\calR$ defined for the beta-function. The only difference is that one of the three channels will need the positions in the four-point function interchanged.

In complete analogy with the standard $\phi^4$ flow, we can make two robust predictions for the $\sigma \chi$ flow. First,
  the dimension of $\chi$ is not renormalized, being controlled by a {{non-local}} kinetic term. Second, using the {{non-local}} equation of motion that relates $\sigma$ and $\chi$, we
  conclude that their dimensions must obey in the IR a shadow relation analogous to (\ref{shadowphi}), 
\beq \label{shadowchi}
[\chi] + [\sigma] = d \,.
\eeq
It is easy to check that these predictions hold to leading order in conformal  perturbation theory, for arbitrary $d$.

On the other hand, to find, {\it e.g.}, the anomalous dimension $\gamma_\varepsilon$ of the energy  operator,  we need a careful evaluation of \reef{eq:B}. 
For $d=2$, we were able to prove analytically that $\gamma_\varepsilon$ vanishes at the leading order $O(g_*^2)$, though we expect higher order corrections to be present,
\beq
\gamma_\varepsilon = O(g_*^4) \qquad  \qquad  \qquad (d = 2)   \, .
\eeq
For $d=3$, we found numerically
\beq
\gamma_\varepsilon = 3.3(5)g_*^2 + O(g_*^4) \qquad (d = 3)\, ,
\eeq
which corresponds to a central value of $B = -6.6$ in (\ref{eq:B}).

{\bf Recombination.} 
The computation of the anomalous dimension of the stress tensor deserves
a special discussion. The SRFP+$\chi$ theory consists in the UV of two decoupled sectors. The SRFP is a local theory, with a conserved local stress tensor  $T_{\mu\nu}$, while the non-local $\chi$ sector has no analogous operator. Clearly, the perturbation by $\sigma \cdot \chi$  couples the two sectors. Locality is lost and 
$T_{\mu\nu}$ acquires an anomalous dimension $\gamma_T$ at the LRFP.
While $\gamma_T$ can be computed by the general method {{outlined}} above, a more illuminating strategy
is {{to}} leverage the phenomenon  of multiplet recombination. (For other recent uses of this strategy in CFT see, {\it e.g.}, \cite{Rychkov:2015naa,Skvortsov:2015pea,Giombi:2016hkj,Roumpedakis:2016qcg}.)

The local stress tensor of the SRFP satisfies the conservation equation $\del^\mu T_{\mu \nu}=0$, meaning that some of his descendants are zero --  the stress tensor is the primary of a short
multiplet of the conformal algebra. Unitarity implies that $\gamma_T = 0$ if and only if $\del^\mu T_{\mu \nu}=0$, so the presence of an anomalous dimension at the LRFP  must be accompanied by a  failure of the conservation equation,
\beq
\del^\mu T_{\mu \nu} \propto V_\nu\ne 0 \,.
\eeq
The short $T_{\mu\nu}$ 
 multiplet becomes long by ``eating'' the $V_\nu$ multiplet. The vector operator  $V_\mu$ must exist in the UV theory as well; this  was puzzling in the standard picture. In our picture, we can instead
easily construct it. The unique candidate  is
\beq
V_{\nu} = \sigma (\del_\nu \chi) -\frac{\Delta_\chi}{\Delta_\sigma}(\del_\nu \sigma)\chi\, ,
\eeq
where the relative normalization is fixed by requiring that it {{be}} a conformal primary.
We can then write
\be
\partial^\mu T_{\mu\nu} = b(g)V_\nu = b_1gV_\nu + O(g^2)\,,
\eeq
which implies
\beq
\langle \del^\mu T_{\mu \nu}(x) V_{\rho}(y) \rangle_g \approx b_1 g_* \langle V_{\nu}(x) V_{\rho}(y)\rangle_0\,. \label{eq:geta}
\eeq
The r.h.s.~of \reef{eq:geta} is easy to evaluate, while the l.h.s.~may be found with one insertion of $\calO$ and use of the Ward identity. This fixes ${b_1 = \Delta_\sigma / d}$.
We also have
\beq
\langle \del^\mu T_{\mu \nu}(x) \del^\rho T_{\rho \sigma}(0) \rangle_{g}\approx b_1^2 g_*^2\langle V_{\nu}(x) V_\sigma(0)\rangle_{0}\,.\label{eq:rec2}
\eeq
The l.h.s.~is the standard two-point function of conformal primaries with dimension $d + \gamma_T$, differentiated twice. Dropping higher powers of $g_*$, \reef{eq:rec2} reduces to
\beq
\gamma_T=\frac{2{\rm S}_{d}^2}{c_T}\frac{\Delta_\sigma (d-\Delta_\sigma)}{d^2+d-2} g_*^2+O(g_*^4)\,. \label{eq:gammaT2}
\eeq
For $d=2$,  we have checked that the general approach to anomalous dimensions, based on computing the principal value \reef{eq:B}, yields a result for $\gamma_T$ in agreement with \reef{eq:gammaT2}.

{\bf Duality.} A pithy way to describe our picture is as an {infrared duality} relating the $\phi^4$ and $\sigma\chi$ flows. We claim that both flows end at the same long-range IR fixed point. Interestingly, the non-local equations of motions give analytic control over several important quantities and allow  precise checks of our proposed duality. The most basic
entry of the duality dictionary is the IR identification 
\beq
\phi \leftrightarrow \sigma \,, \qquad \phi^3 \leftrightarrow \chi \,.
\eeq
Combining the non-renormalization of the dimensions of  $\phi$ and of $\chi$ with the shadow relations (\ref{shadowphi}), (\ref{shadowchi}), one easily
checks that the dimensions of the dual pairs agree at the IR fixed point  \cite{Caveat},
\beq
[\phi] = [\sigma]=\frac{d-s}{2}\,,\qquad [\phi^3]=[\chi]= \frac{d+s}{2}  \, .\label{eq:duality} 
\eeq 
The non-local equations of motion can also be used to relate OPE coefficients involving shadow pairs of operators. Let ${\cal O}_1$ and ${\cal O}_2$ 
denote two arbitrary scalar primary operators and $\lambda_{12 \tilde \phi}$ (respectively  $\lambda_{12 \tilde \phi^3}$) their three-point coupling with the unit-normalized operator $\tilde \phi$ (respectively $\tilde \phi^3$).   It was shown in \cite{LRI} that the ratio
$\lambda_{12 \tilde \phi^3}/\lambda_{12 \tilde \phi}$  is given by a universal formula that depends only on $d$, on the operator dimensions, 
and on the normalization of  $\phi$. An analogous reasoning in the dual flow leads to a similar formula for $\lambda_{12 \tilde \chi}/\lambda_{12 \tilde \sigma}$, which must in fact agree with $\lambda_{12 \tilde \phi^3}/\lambda_{12 \tilde \phi}$ if the duality is to hold. Remarkably, it does, but only if the normalizations of $\phi$ and $\chi$ are related in a precise way. Defining
\beq
\langle \phi(x) \phi(0) \rangle = \frac{1 + \rho(\epsilon)}{|x|^{2\Delta_\phi}}\,, \quad \langle \chi(x) \chi(0) \rangle  =  \frac{1 + \kappa (\delta)   }{|x|^{2\Delta_\chi}}\, , 
\eeq
compatibility of the OPE ratios demands
\beq
\frac{\kappa(\delta) }{1 + \kappa(\delta)} =  \frac{1 + \rho(\epsilon)}{\rho(\epsilon)}\,.
\eeq
Since $\kappa(\delta) = O(\delta)$ for $\delta \to 0$, 
this predicts that the normalization of $\phi$ vanishes as we approach the {{crossover}}, where the description in terms of the $\sigma \chi$ flow becomes
weakly coupled. Conversely, since
$\rho (\epsilon) = O(\epsilon^2)$,  the normalization of $\chi$ must vanish for $\epsilon \to 0$, 
 where  the $\phi^4$ flow becomes weakly coupled. 
 That the  normalization of $\phi$ should vanish at the continuous {{crossover}} has been previously noticed in \cite{LRI}, and it was also argued in \cite{Parisi} using a large $N$ expansion. 

{\bf Discussion.} In this Letter we have  put forward a new theory for the long-range to short-range crossover. Prior to our work, the understanding of this crossover was incomplete at best. Crucially, our new qualitative picture allowed us to greatly advance the quantitative side of the story, hitherto nonexistent. We obtained a number of predictions for the critical exponents near the crossover, which in principle can be confirmed by Monte-Carlo simulations and perhaps even experiments. Also, the existence of $\chi$ leads to experimentally verifiable predictions even in the short-range regime $\mysigma>\mysigma_*$, where it is decoupled. The point is that it is decoupled from the short-range fixed point fields, but not from the lattice operators, so it should be possible to detect $\chi$ via lattice measurements.

 While we have focused on the long-range {Ising} model, it's clear that most of the learned lessons apply to the $O(N)$ case as well. Still more generally, our $\sigma\chi$-flow construction can be used 
with any CFT in place of the SRFP. Just pick a scalar CFT operator, call it $\sigma$ again, of dimension $\Delta$, and couple it to a non-local gaussian field $\chi$ of dimension $d-\Delta-\delta$, $\delta\ll1$. If the quantum correction to the beta-function has the right sign, we will then obtain a continuous family (parametrized by $\delta$) of non-local conformally invariant theories which are deformations of the original local CFT. It will be unitary if the original CFT was unitary and if $\chi$ is above the unitarity bound. One interesting application of this idea is to the long-range Potts model. We find it likely that explorations along these lines
will lead to other examples of non-local IR dualities.

\vspace{10pt}

\begin{acknowledgments}
We are grateful to the Galileo Galilei Institute of Theoretical Physics where this work was initiated. SR is supported by Mitsubishi Heavy Industries as an ENS-MHI Chair holder. SR and BZ are supported by the National Centre of Competence in Research SwissMAP funded by the Swiss National Science Foundation. LR and SR are supported by the Simons Foundation grants 397411 and 488655 (Simons collaboration on the Non-perturbative bootstrap). CB is supported by the Natural Sciences and Engineering Research Council of Canada.

\end{acknowledgments}


\begin{thebibliography}{99}

\bibitem{note1} Formally our considerations apply also to non-integer
  dimensions in the range $1<d<4$.
  
 \bibitem{note2} We reserve the letter $\sigma $ for the short-range Ising spin field. What we have called $s$ is usually denoted $\sigma $.
  
 \bibitem{Note101} The action (\ref{standardflow}) is appropriate for the interval $0<s<2$, which
  includes the long-range to short-range crossover point. Beyond $s=2$, we would have to add the
  local kinetic term $(\partial \phi )^2$ which becomes relevant in this regime.
  
 \bibitem{note3} Throughout the paper scaling dimensions of various fields
  $X$ are denoted interchangeably by $\Delta _X$ or $[X]$.
  
  \bibitem{Note100}
  The long-distance behavior is again mean-field with powerlike (as opposed to exponential) correlations when we are away from the transition, in the disordered phase. Nonalyticity of the $\phi$ propagator in momentum space, $1/(|p|^\mysigma+t)$, at $p=0$ is what causes this powerlike falloff. The IR value of the $\phi$ dimension can be read off as $[\phi]_{\rm dis}=d-[\phi]_{\rm UV}$. This also holds for fluctuations of $\phi$ around the mean value in the ordered phase.

\bibitem{Sak} J. Sak, Phys. Rev. B {\bf 8}, 281 (1973).

\bibitem{Fisher} M. E. Fisher, S.-k. Ma, and B. Nickel, Phys.Rev.Lett. {\bf 29}, 917 (1972).

\bibitem{Sak2} J. Sak, Phys. Rev. B {\bf 15}, 4344 (1977).



\bibitem{Cardy} J. L. Cardy, {\it Scaling and renormalization in statistical physics} (Cambridge University Press, Cambridge, UK, 1996), section 4.3.

\bibitem{LRI} M. F. Paulos, S. Rychkov, B. C. van Rees, and B. Zan, Nucl. Phys. {\bf B902}, 246 (2016), arXiv:1509.00008 [hep-th].

\bibitem{noteRecomb} The puzzle can be stated more formally in terms of ``recombination rules''  of unitary representations of the conformal algebra ${\mathfrak so}(d+1, 1)$.  (The enhancement of scale invariance to conformal invariance at the long-range fixed point --
 even in the absence of a local stress tensor -- has been recently demonstrated 
 in \cite{LRI}.)
The standard  stress tensor of the SRFP is the lowest weight (conformal primary) state of the {\it shortened} spin-two representation ${\cal C}_{\ell = 2}^{d}$,
while the non-conserved spin-two operator of the LRFP is the conformal primary of 
the {\it long} spin-two representation 
 ${\cal A}_{\ell = 2}^\Delta$, where unitarity demands $\Delta \geq d$.
When the unitarity bound is saturated, the long spin-two representation decomposes into the following semi-direct sum: 
${\cal A}_{\ell = 2}^d  \simeq {\cal C}_{\ell = 2}^{d} \oplus {\cal A}_{\ell = 1}^{d+1}$.  In other terms, the shortened spin-two representation can only become long by recombining
with ${\cal A}_{\ell = 1}^{d+1}$, an additional spin-one representation whose conformal primary
 $V_\mu$ is, however, missing in the SRFP.

\bibitem{noteSign} Note that for real $g$, the $\calO_{\rm Sak}$ generated this way has ferromagnetic (negative) sign, as it should.

\bibitem{noteCoincides} Note that $[\chi]$ coincides with $[\phi]_{\rm dis}$, the dimension of $\phi$ in the disordered phase of the long-range model \cite{Note100}, suggesting a similarity between the short-range critical fluctuations on the long-range correlations and the disordering effects of temperature.

\bibitem{LRI2} C. Behan, L. Rastelli, S. Rychkov, and B. Zan, (2017), arXiv:1703.05325 [hep-th].

\bibitem{Zamolodchikov:1987ti} 
  A.~B.~Zamolodchikov,
  Sov.\ J.\ Nucl.\ Phys.\  {\bf 46}, 1090 (1987)
  [Yad.\ Fiz.\  {\bf 46}, 1819 (1987)].

\bibitem{Brezin:1990sk} 
 J.~L.~Cardy, {\it Conformal Invariance and Statistical Mechanics}, in E.~Brezin and J.~Zinn-Justin, Eds., Proceedings, 49th Session of the Les Houches Summer School in Theoretical Physics,
  Amsterdam, Netherlands: North-Holland (1990) 640 p
  
\bibitem{Cappelli:1991ke} 
  A.~Cappelli, J.~I.~Latorre and X.~Vilasis-Cardona,
  Nucl.\ Phys.\ B {\bf 376}, 510 (1992)
    [hep-th/9109041].

\bibitem{noteDiv} The divergence is in fact observable-independent and $\calO(\infty)$ is just a convenient choice.

\bibitem{Zohar} Z. Komargodski and D. Simmons-Duffin, J.\ Phys.\ A: Math.\ Theor.\ {\bf 101}, 1053 (2016), arXiv:1603.04444 [hep-th].



\bibitem{BPZ} A. Belavin, A. M. Polyakov, and A. Zamolodchikov, Nucl.Phys. {\bf B241}, 333 (1984).

\bibitem{c-min} S. El-Showk, M. F. Paulos, D. Poland, S. Rychkov, D. Simmons-Duffin, and A. Vichi, J.Stat.Phys. {\bf 157}, 869 (2014),
arXiv:1403.4545 [hep-th].

\bibitem{lightcone} D. Simmons-Duffin, (2016), arXiv:1612.08471 [hep-th].



\bibitem{noteErrors} For $d=2$, the quoted error comes from estimating  the error in the numerical integration. For $d=3$, the main source of uncertainty is the
  approximate knowledge of the 3d Ising spectrum and OPE coefficients. See \cite{LRI2} for a detailed analysis of the error estimate.  






\bibitem{Rychkov:2015naa} S. Rychkov, and Z. M. Tan, J.Phys. {\bf A48}, 29FT01 (2015), arXiv:1505.00963 [hep-th].

\bibitem{Skvortsov:2015pea} E.~D.~Skvortsov, in: {\it Proceedings, International Workshop on Higher Spin Gauge
                        Theories: Singapore, November 4-6, 2015}, 103-137 (2017),  
    arXiv:1512.05994 [hep-th].

\bibitem{Giombi:2016hkj} S. Giombi, and V. Kirilin, JHEP {\bf 11}, 068 (2016), arXiv:1601.01310 [hep-th].

\bibitem{Roumpedakis:2016qcg} K. Roumpedakis, arXiv:1612.08115 [hep-th].



\bibitem{Caveat} While the non-renormalization theorem and the 
the shadow relations have only be  proven to all orders in conformal perturbation theory \cite{LRI,LRI2}, the overall consistency of our picture leaves
 little doubt that they hold exactly. A non-perturbative proof of the non-renormalization
of $[\phi]$ has been suggested to us by P. Liendo and M. Meineri, see footnote 25 of \cite{LRI2}.

\bibitem{Parisi} E. Brezin, G. Parisi, and F. Ricci-Tersenghi, J.Stat.Phys. {\bf 154}, 855 (2014), arXiv:1407.3358 [cond-mat.stat-mech].

\end{thebibliography}
\end{document}